\documentclass[printer]{aa} % for a referee version
\usepackage{graphicx}
\usepackage{caption}
\usepackage{subcaption}
\usepackage{epstopdf}
\usepackage{amsmath}
\usepackage[english]{babel}
\usepackage{amssymb}
\usepackage{wrapfig}
\usepackage{pdflscape}
\usepackage{lscape}
\usepackage{longtable}
\usepackage{appendix}
\usepackage{textcomp}
\usepackage{multirow}
\bibpunct{(}{)}{;}{a}{}{,}
\usepackage{array}
\usepackage{ragged2e}
\usepackage{adjustbox}
\usepackage{txfonts}
\usepackage{xcolor}
\usepackage{float}
\usepackage[bookmarksnumbered=true]{hyperref} 
\hypersetup{
    colorlinks = true,
    linkcolor = blue,
    anchorcolor = blue,
    citecolor = blue,
    filecolor = blue,
    urlcolor = blue
    }

\begin{document}

   \title{A host galaxy study of southern narrow-line Seyfert 1 galaxies}
 \titlerunning{NLS1 host morphology}

   \author{I. Varglund\inst{1,2}\thanks{irene.varglund@aalto.fi}
          \and
          E. Järvelä\inst{3,4}\thanks{Dodge Family Prize Fellow in The University of Oklahoma}
          \and 
          S. Ciroi\inst{5}
          \and
          M. Berton\inst{6}
          \and
          E. Congiu\inst{6}
          \and
          A. Lähteenmäki\inst{1,2}
          \and
          F. Di Mille\inst{7}
          }

    \institute{Aalto University Metsähovi Radio Observatory, Metsähovintie 114, FI-02540 Kylmälä, Finland
         \and
             Aalto University Department of Electronics and Nanoengineering, P.O. Box 15500, FI-00076 AALTO, Finland
             \and
             European Space Agency (ESA), European Space Astronomy Centre (ESAC), Camino Bajo del Castillo s/n, 28692 Villanueva de la Cañada, Madrid, Spain
             \and
             Homer L. Dodge Department of Physics and Astronomy, The University of Oklahoma, 440 W. Brooks St., Norman, OK 73019, USA
             \and
            Dipartimento di Fisica e Astronomia "G. Galilei", Università di Padova, Vicolo dell’Osservatorio 3, 35122 Padova, Italy   
         \and
             European Southern Observatory (ESO), Alonso de C\'ordova 3107, Casilla 19, Santiago 19001, Chile
%        \and
%             Finnish Centre for Astronomy with ESO (FINCA), University of Turku, Vesilinnantie 5, FI-20014 University of Turku, Finland
        \and
            Las Campanas Observatory, Carnegie Observatories, Colina El Pino, Casilla 601, La Serena, Chile}

   \date{Received }

  \abstract
{We studied seven nearby narrow-line Seyfert 1 (NLS1) galaxies in $J$- and $Ks$-bands with redshifts varying from 0.019 to 0.092. This is the first multi-source study targeting the hosts of southern NLS1 galaxies. Our data was obtained with the FourStar instrument of the 6.5~m Magellan Baade telescope at the Las Campanas Observatory (Chile). The aim of our study is to determine the host galaxy morphologies of these sources by using GALFIT. We were able to model six out of the seven sources reliably. Our conclusion is that all of the reliably modelled sources are disk-like galaxies, either spirals or lenticulars. None of these sources present elliptical morphology. Our findings are in agreement with the hypothesis that disk-like galaxies are the main host of jetted NLS1 galaxies. Taking advantage of observations in two bands, we also produced a $J - Ks$ colour map of each source. Five of the six colour maps show significant dust extinction near the core of the galaxy -- a feature often seen in gamma-ray-detected jetted NLS1 galaxies, and interpreted to be a consequence of a past minor merger.}

  \keywords{galaxies: active -- galaxies: Seyfert -- galaxies: structure -- infrared: galaxies}

   \maketitle
%
%-------------------------------------------------------------------

\section{Introduction}
\label{sec:intro}

\begin{table*}
\caption[]{Basic properties of the sample and the observations. }
\centering
\scalebox{0.9}{
\begin{tabular}{l l l l l l l l l l}
\hline\hline
Source      & Short name             & $z$     & Scale   & RA          & Dec         & Exp ($J$) & Exp ($Ks$) & Date    & Seeing\\
                         &    &    & (kpc$/$\arcsec    ) & (J2000.0)   & (J2000.0)   & (s)   & (s)    &  (year/month/day)              & (\arcsec    )  \\ \hline
6dFGS J0622335-231742 & J0622-2317 & 0.019 &  0.370 & 06:22:34 & -23:17:42 & 960 & 384 & 2020/02/08 & 0.7 \\
6dFGS J0952191-013644 & J0952-0136 & 0.037 & 0.705 & 09:52:19 & -01:36:44 & 583 & 384 & 2020/02/08 & 1.0 \\
6dFGS J1511598-211902 & J1511-2119 & 0.044 & 0.832 & 15:12:00 & -21:19:02 & 640 & 384 & 2019/07/19 & 0.8 \\
6dFGS J1522287-064441 & J1522-0644 & 0.083 & 1.500 & 15:22:29 & -06:44:41 & 640 & 384 & 2019/07/20 & 0.7 \\
6dFGS J1615191-093613 & J1615-0936 & 0.064 & 1.182 & 16:15:19 & -09:36:13 & 640 & 384 & 2019/07/20 & 0.7 \\
6dFGS J1628484-030408 & J1628-0304 & 0.092 & 1.645 & 16:28:48 & -03:04:08 & 640 & 384 & 2019/07/20 & 0.6 \\
6dFGS J1646104-112404 & J1646-1124 & 0.074 & 1.351 & 16:46:14 & -11:24:04 & 872 & 523 & 2019/07/20 & 0.7 \\
   
\hline
\end{tabular}
}
\tablefoot{Columns: (1) Source name, (2) Short name of the source, (3) Redshift, (4) Scale at the redshift of the source, (5) Right ascension, (6) Declination, (7) Total exposure time in $J$-band, (8) Total exposure time in $Ks$-band, (9) Observation dates, (10) Average seeing during the observations measured from the images in the band used for modelling.}
\label{tab:sample}
\end{table*}

% Perusjutut
Narrow-line Seyfert 1 (NLS1) galaxies \citep{1985osterbrock1} are a class of peculiar active galactic nuclei (AGN). The full width at half maximum (FWHM) of H$\beta$ is by definition less than 2000~km s$^{-1}$ \citep[]{1989goodrich1}. This class is often characterised by its unique optical spectral line features. One of the most common identifiers of NLS1 galaxies is the faint [O~\textsc{III}] emission with respect to H$\beta$ ($\rm S([O~\textsc{III}]\lambda5007) / S(H\beta) < 3$, \citealt{1985osterbrock1}). Though not all NLS1 galaxies, a considerable fraction of them exhibit strong Fe~\textsc{II} multiplets \citep[]{1992boroson}. Some NLS1 galaxies show blueshifted line profiles in high-ionization lines, indicating outflows \citep[e.g.][]{2002zamanov,2005boroson,2008komossa,2021berton2}. 

% BH-mass ja unevolved
NLS1 galaxies are intrinsically different from broad-line Seyfert (BLS1) galaxies, for example, in their black hole masses, Eddington ratios, and their multiwavelength and variability properties. The general consensus is that, unlike BLS1 galaxies, NLS1 galaxies are powered by low-mass central supermassive black holes ($M_{\text{BH}} \sim 10^{6} M_{\sun}$ - $10^{8} M_{\sun}$; \citealp[e.g.][]{2011peterson, 2015jarvela, 2016cracco1, 2018chen}). This has been confirmed, for example, by reverberation mapping campaigns \citep[e.g.][]{2016wang, 2018du, 2019du1}. Due to the low black hole mass, NLS1 galaxies are thought to be in an early evolutionary stage with respect to BLS1 galaxies and other classes of AGN \citep{2000mathur}. They also exhibit high Eddington ratios, generally between 0.1 and 1, but in some cases also exceeding unity \citep{1992boroson, 2018marziani, 2022tortosa}. 

% Radio
NLS1 galaxies have previously been thought to not have significant radio emission \citep{2006komossa1,2015jarvela}. Of the currently identified NLS1 galaxies, roughly one sixth, $\sim16\%$, have been detected at radio frequencies: $\sim6\%$ are weak radio emitters, and $\sim10\%$ have significant radio emission \citep{2006komossa1}. The rest of the NLS1 galaxies, $\sim84\%$, have not been detected at radio frequencies, yet.

% Jetillisyys
Finding disk galaxies, such as BLS1 galaxies, with jets is not uncommon \citep{2006keel, 2014bagchi, 2022wu, 2023gao}, however, these sources in general have stellar masses similar to those of elliptical galaxies and are thus atypical among disk-like galaxies. NLS1 galaxies, on the other hand, have pseudo-bulges and low S\'{e}rsic indices, indicating that their stellar masses are low \citep{2020sanchez}. Thus the discovery of relativistic jets in some NLS1 galaxies was unexpected. These jets appear similar, but less powerful, to those in blazars \citep[e.g.][]{2006komossa1, 2006zhou,  2008yuan, 2011foschini, 2015foschini1, 2017lahteenmaki1, 2018lahteenmaki1}. Typically, relativistic jets manifest strongly in radio and dominate the whole band. However, also examples of jetted sources with extremely weak low-frequency ($<$ 10~GHz) radio emission have been found \citep[][Järvelä et al. in prep.]{2018jarvelaphd,2020berton, 2021jarvela, 2021sbarrato}, questioning the utilisation of the emission in this regime as a proxy for jettedness. The opposite has also been observed, when sources with significant radio emission at low frequencies do not host relativistic jets, even though that would be expected \citep{2022jarvela}. This anomaly is caused by the presence of significant star formation activity \citep{2015caccianiga,2019ganci, 2020berton}. 

% Earlier studies and morphology, spiral
There are several individual source analysis studies \citep{2008anton, 2014leontavares, 2016kotilainen, 2017dammando, 2017olguiglesias, 2018dammando, 2019berton, 2021hamilton, 2022vietri}, as well as numerous large sample studies of the hosts of NLS1 nuclei \citep{2001krongold, 2003crenshaw, 2007ohta1, 2011orban, 2012mathur, 2014leontavares, 2017jarvela, 2020olguiglesias, 2022varglund}. Nearly all of these studies have identified them as disk-like. Elliptical hosts have been an anomaly, however, some cases seem to exist \citep{2017dammando, 2018dammando}.

In this paper we examined the hosts of seven southern NLS1 galaxies that are also part of an optical spectral study and a multi-frequency radio imaging study. This is the first systematic host galaxy investigation of southern NLS1 galaxies, and will be expanded in the future to allow statistically significant comparison of their radio and optical properties. Our goal was to determine the morphology of the hosts to improve the statistical weight of the recent findings that most NLS1 nuclei are hosted in disk-like galaxies. Furthermore, we also studied the colour maps as they could tell us more about the properties of the core, such as its dust content, and shed light on possible merger history. Ultimately, NLS1 galaxies as early-stage AGN, and conveniently residing in the local Universe, can elucidate the evolution of AGN over cosmic time.

Throughout this paper we assume standard $\lambda$CDM cosmology with parameters H$_{0}$ = 73 km s$^{-1}$ Mpc$^{-1}$, $\Omega_{\text{matter}}$ = 0.27, and $\Omega_{\text{vacuum}}$ = 0.73 \citep{2007spergel1}.

\begin{table}
\caption[]{Radio properties of the sample (F = flat, S = steep), and the black hole mass estimates. }
\centering
\begin{tabular}{llllll}
\hline\hline
Source        & $S_{\mathrm{1.4~GHz}}$ & $S_{\mathrm{5.5~GHz}}$ & log $M_{\text{BH}}$ & &  \\
J0622-2317  & 4.3  & 0.9    & 6.47 & F  & Diffuse \\
J0952-0136  & 59.8 & 21.3   & 6.33 & S & Jetted \\
J1511-2119  & 46.9 & 18.2   & 6.63 & S & Jetted? \\
J1522-0644  & 14.4 & 5.5    & 6.66 & S & Compact \\
J1615-0936  &      & 0.7    & 6.94 & S & Jetted? \\
J1628-0304  &      & 0.1    & 6.72 & F & Compact \\
J1646-1124  & 38.3 & 10.8   & 7.07 & S & Compact \\
\hline
\end{tabular}
\tablefoot{All values and radio properties from \citet{2020chen}.} \label{tab:radio}
\tablefoot{Columns: (1) Source name, (2) 1.4 GHz FIRST maximum flux density level, (3) 5.5 GHz maximum flux density, (4) Black hole mass, (5) Radio spectrum, and (6) Radio morphology}
\end{table}
%--------------------------------------------------------------------
%\section{Sample selection}
%\label{sample}

\section{Observations and data reduction}
\label{sec:obs}

Our sample of seven sources was selected from \citet{2018chen}. Our selection criteria were based on the sources' radio properties and the possible presence of jets \citep{2020chen}. The sources chosen for this study have all been detected at 5.5~GHz, with five of them also having been detected at 1.4~GHz. We have presented the radio flux densities as well as some basic characteristics of our earlier observations in Table~\ref{tab:radio}. The radio spectrum of five of the sources is steep, while the final two sources have a flatter spectrum. Based on the results of \citet{2020chen}, J0952-0136, J1511-2119, and J1615-0936 were expected to harbour relativistic jets. J0952-0136 has been identified as a jetted source also by \citet{2015doi}. To investigate the radio properties of the \citet{2020chen} sample in more detail, we have obtained JVLA observations in 15, 22, and 33~GHz (Berton et al. in prep). All of our sources have a low redshift ($<0.1$) and were observed with FourStar, mounted on the 6.5~m Magellan Telescope. Five sources were observed in July, 2019, while the rest were observed in February, 2020. One source, J0952-0136, was saturated and could not be modelled.

The field-of-view (FoV) of FourStar is 5.4' $\times$ 5.4', with a scale of 0.159\arcsec px$^{-1}$. All of our sources were observed in both $J$- and $Ks$-bands. When observing in $J$-band, we used 4-point dithering for all sources but J0622-2317 for which we used 6-point dithering due to its brightness. In the $Ks$-band the dithering used was 6-point for all sources. 

%Data redusointi, tynkästi kirjoitettu
For data reduction, we used IRAF. We first created a flat field image for each source for each respective band, to remove eventual signatures in the data produced by the instrument. This flat was applied to each dithered image. After creating the flat field, we combined the dithered images to produce one image per each dithering sequence. Then we continued by creating a superflat to remove the final effects of the background. As the result, the mean sky value is essentially zero. To finalise the reduction, we aligned the created images and combined them. The data reduction process was done for both filters. After producing the final images, we determined the zero point magnitude for each source in both bands using 2MASS\footnote{https://www.ipac.caltech.edu/2mass} magnitudes.

\section{Data analysis}
\label{data}

To perform the photometric decomposition of the host we used the band with better seeing (see Table~\ref{tab:sample}), which in all cases of this study resulted to be the $J$-band. $Ks$-band images were only used to create the colour maps. Our goal was to determine the host galaxy morphology and the related model parameters that best describe the source. 

To perform the photometric decomposition, we followed the same procedure as described in \citet{2022varglund} with a few exceptions. The PSF model was built primarily with DAOPHOT and secondarily through choosing a star from the image. No separate PSF stars were observed for this project. Furthermore, our cutout image sizes were either $\sim$ 100 $\times$ 100~px or $\sim$ 200 $\times$ 200~px. The size of the image was dependent on the cutout size around the host galaxy we were modelling. The sky background parameter was estimated through two methods: the method described in \citet{2022varglund} and by using a large annulus that encompassed roughly a fifth of our image. We made sure that no source was inside the annulus, before determining the sky background value. The former method was used to confirm the results of the latter method.  Due to the used data reduction process, the sky background value was virtually zero. 

The fitting process with GALFIT version 3 \citep{2010peng1}, creating the radial surface plot with the IRAF \citep{1987stetson} task $ELLIPSE$, and the error estimation method was identical to the process in \citet{2022varglund}. The radial surface plot was created to study both the accuracy of the fit as well as to see how the brightness of the components changes as a function of the distance from the core. On average the limiting surface brightness of the images is around 22-23 mag arcsec$^{-2}$ whereas all of our sources are considerably brighter than the limit, and thus we can model them reliably up to large radii. 

In the case of a suspected bar-component based on the best-fit parameters we used an additional diagnostic to determine if the component truly is a bar. To do this, we extracted the position angle (PA) and ellipticity parameters of the isophotes obtained from the \textit{ELLIPSE} output of the galaxy image. These parameters were plotted against the radius. The plot indicates the presence of a bar if there is a region in which the ellipticity increases while the PA remains mostly constant ($\Delta$PA $< 20$\textdegree) \citep{1995wozniak1, 2007mendezdelmestre1}.

Finally, since we had images from two bands for each source, we were able to create colour maps of all of our sources. To obtain the colour map, we used the equation below: 

\begin{equation}
    \texttt{Colour image pixel value} = (\text{zp}_J -  \text{zp}_{Ks}) - 2.5\Bigg[\text{log}_{10}\bigg(\frac{J}{Ks}\bigg)\Bigg] ,
\end{equation}

The data used when calculating the colour image was corrected for the different exposure times of the bands. The images were aligned by using \textit{GEOMAP}, \textit{GEOTRAN}, and \textit{IMALIGN} to guarantee that the images were perfectly aligned. The data processing allowed for the PSFs to be nearly identical, improving the colour maps. In the above equation, $zp$ stands for the zero point magnitude, with the subscript defining the band in question. The contour levels in the maps are defined using the Ks- band image sky standard deviation (std) and are at 20, 40, 60, 80, 100, 180, 360, 720, and 1440 $\times$ std.

\begin{table*}
\caption[]{Best-fit parameters of J0622-2317. $\chi^2_{\nu}$ = 0.988 $\substack{+0.00 \\ -0.00}$.}
\centering
\begin{tabular}{l l l l l l l}
\hline\hline
Function & Mag & $r_\text{e}$ & $n$ & Axial & PA & Notes \\
 & & (kpc) & & ratio & (\textdegree) & \\ 
\hline
PSF 1 & 14.70 $\substack{+0.01 \\ -0.01}$ & & & & & \\
S\'{e}rsic 1 & 11.98 $\substack{+0.01 \\ -0.02}$ & 3.20 $\substack{+0.19 \\ -0.00}$ & 1.65 $\substack{+0.01 \\ -0.00}$ & 0.80 $\substack{+0.00 \\ -0.00}$ & 41.91  $\substack{+0.01 \\ -0.02}$ & Disk and bulge \\
%PSF 2 & 13.80 $\substack{+0.01 \\ -0.01}$ & & & & &  Nearby source \\
%PSF 3 & 17.54 $\substack{+0.01 \\ -0.01}$ & & & & &  Nearby source \\
%PSF 4 & 18.60 $\substack{+0.01 \\ -0.01}$ & & & & &  Nearby source \\
\hline   % centered "correctly"
\end{tabular}
\tablefoot{Columns: (1) Function used for modelling, (2) Magnitude, (3) Effective radius, (4) S\'{e}rsic index, (5) Axial ratio, (6) Position angle, (7) Additional notes on the sources.}
\label{tab:j0622}
\end{table*}

\begin{figure*}
\centering
\adjustbox{valign=t}{\begin{minipage}{0.34\textwidth}
\centering
\includegraphics[width=1\textwidth]{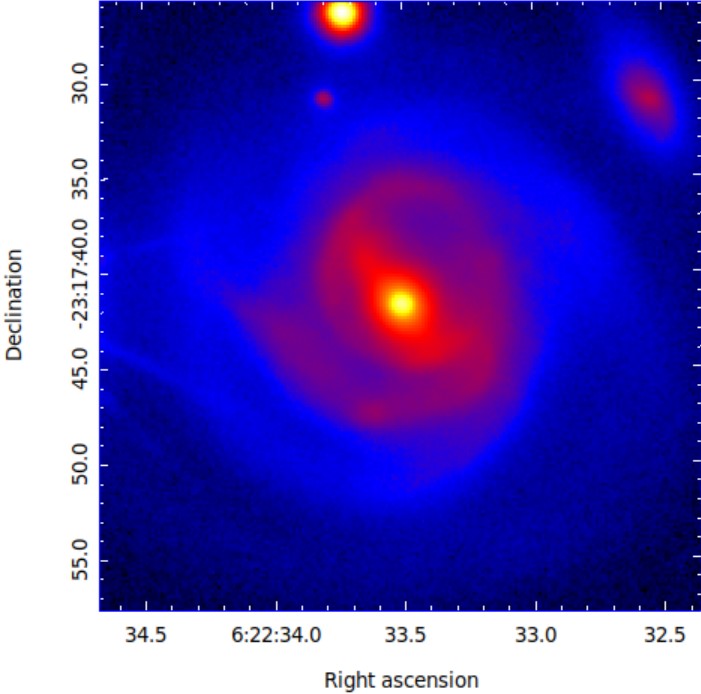}
\end{minipage}}
\adjustbox{valign=t}{\begin{minipage}{0.31\textwidth}
\centering
\includegraphics[width=0.95\textwidth]{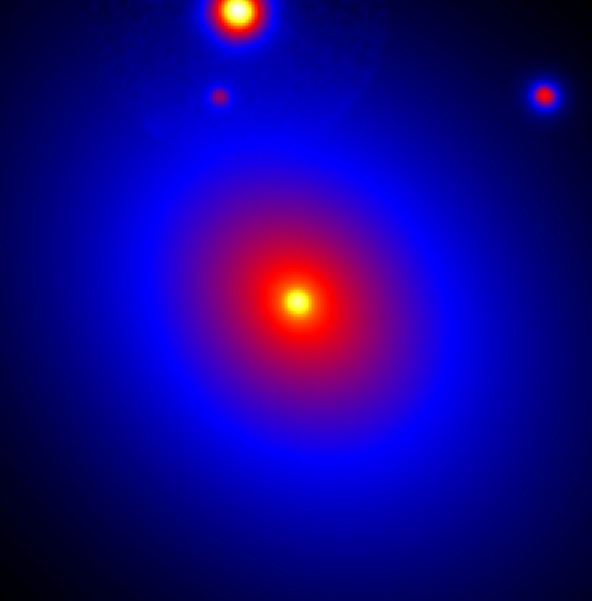}
\end{minipage}}
\adjustbox{valign=t}{\begin{minipage}{0.31\textwidth}
\centering
\includegraphics[width=0.95\textwidth]{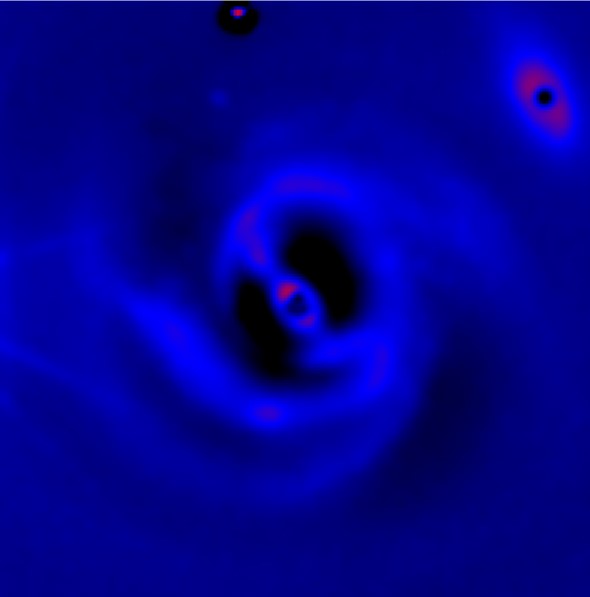}
\end{minipage}}
\hfill
    \caption{\textit{J}-band images of J0622-2317. The FoV is 31.8\arcsec    $/$ 11.8~kpc in all images. \emph{Left panel:} observed image, \emph{middle panel:} model image, and \emph{right panel:} residual image, smoothed over 3px.}  \label{fig:j0622}

\adjustbox{valign=t}{\begin{minipage}{0.49\textwidth}
\centering
\includegraphics[width=0.95\textwidth]{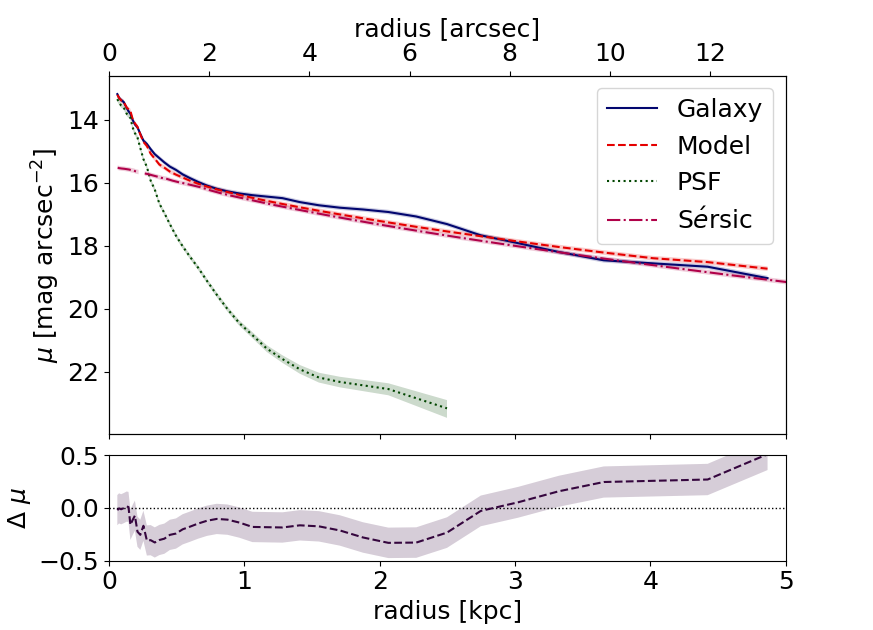}
\caption{Radial surface brightness profile plot of J0622-2317. The blue line represents the galaxy component, the dashed red line shows the model, the dotted green line marks the PSF, and finally the dashed pink line shows the S\'{e}rsic component. The error of each component is shown by the shaded area surrounding each curve. }
\label{fig:j0622comps}
\end{minipage}}
\hfill
\adjustbox{valign=t}{\begin{minipage}{0.49\textwidth}
\centering
\includegraphics[width=0.95\textwidth]{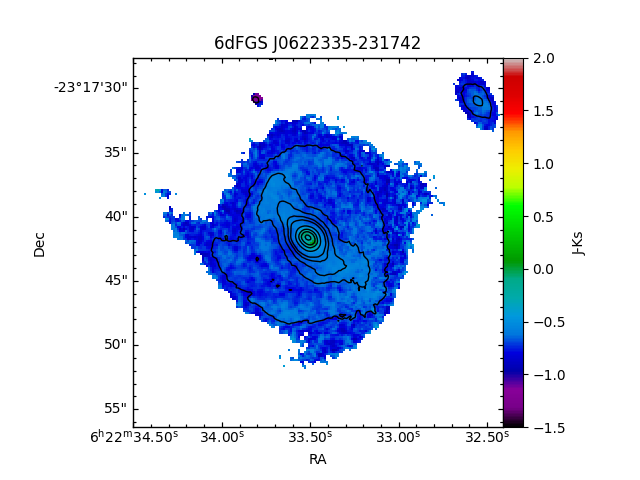}
\caption{\textit{$J$ - $Ks$} colour map of J0622-2317.}
\label{fig:j0622color}
\end{minipage}}
\hfill
\end{figure*}

\section{Individual source analysis}
\label{sec:individual}

\subsection{J0622-2317}

According to \citet{2020chen} this source has a flat in-band radio spectrum, and a diffuse radio morphology. Additional information of this source and all the sources of our sample can be found in Table~\ref{tab:radio}. A visual inspection shows a very clear spiral structure and even hints at the existence of a bar (Fig.~\ref{fig:j0622}).

The size of the fitting region was limited due to a nearby, very bright, saturated star. The best-fit results can be seen in Table~\ref{tab:j0622}. In addition to the NLS1, also three nearby sources were fitted using the PSF model to obtain a more realistic $\chi^2_{\nu}$ for the FoV. We have most likely fitted a disk, or a combination of a disk and a pseudo-bulge component that cannot be separated and modelled individually. The bulge may be unresolved, but its light is still present in the image, thus affecting the model, or the bulge is either not bright enough or does not have different enough properties compared to the disk. The component is small, however, the S\'{e}rsic index is high for a disk so it is possible that a bulge component is included. The intensity plots of the source can be seen in Fig.~\ref{fig:j0622}. There are clear residuals left due to the spiral arms and some slight residuals near the core of the galaxy. We tried to include a bar into the model, but were unable to obtain good results. 

%It might be that the bar-like feature is due to the spiral arms. 

The radial surface plot can be seen in Fig~\ref{fig:j0622comps}. As expected, the model image curve does not perfectly follow the curve for the galaxy image. This is most likely due to the spiral arms not being completely modelled. The colour map can be seen in Fig.~\ref{fig:j0622color}. On average, Seyfert galaxies have a \textit{$J$ - $Ks$} magnitude between 0 and 2.5, and when the magnitude difference is 1.3 or more, a galaxy is considered red \citep{2000jarrett}. Unlike typical Seyfert galaxies, J0622-2317 is blue. However, the magnitude of the core of the galaxy is above 0, and is thus within a common range found in NLS1 galaxies. As with the other images, the spiral arms are clearly visible in the colour map. 

\subsection{J1511-2119}

\begin{table*}
\caption[]{Best-fit parameters of J1511-2119. $\chi^2_{\nu}$ = 0.916 $\substack{+0.00 \\ -0.02}$. }
\centering
\begin{tabular}{l l l l l l l}
\hline\hline
Function & Mag & $r_\text{e}$ & $n$ & Axial & PA & Notes \\
 & & (kpc) & & ratio & (\textdegree) & \\ 
\hline
PSF 1 & 14.24 $\substack{+0.03 \\ -0.03}$ & & & & & \\
S\'{e}rsic & 15.90 $\substack{+0.03 \\ -0.03}$ & 3.93 $\substack{+0.15 \\ -0.00}$ & 2.96 $\substack{+0.08 \\ -0.00}$ & 0.90 $\substack{+0.00 \\ -0.00}$ & -70.82  $\substack{+0.00 \\ -0.12}$ & Bulge and disk \\
%PSF 2 & 16.09 $\substack{+0.03 \\ -0.03}$ & & & & &  Nearby source \\
%PSF 3 & 20.72 $\substack{+0.03 \\ -0.03}$ & & & & &  Nearby source \\
%PSF 4 & 19.14 $\substack{+0.03 \\ -0.03}$ & & & & &  Nearby source \\
%PSF 5 & 19.19 $\substack{+0.03 \\ -0.03}$ & & & & &  Nearby source \\
%PSF 6 & 19.65 $\substack{+0.03 \\ -0.03}$ & & & & &  Nearby source \\
%PSF 7 & 17.62 $\substack{+0.03 \\ -0.03}$ & & & & &  Nearby source \\
%PSF 8 & 18.15 $\substack{+0.03 \\ -0.03}$ & & & & &  Nearby source \\
%PSF 9 & 18.59 $\substack{+0.03 \\ -0.03}$ & & & & &  Nearby source \\
%PSF 10 & 17.71 $\substack{+0.03 \\ -0.03}$ & & & & &  Nearby source \\
%PSF 11 & 17.83 $\substack{+0.03 \\ -0.03}$ & & & & &  Nearby source \\
\hline   % centered "correctly"
\end{tabular}
\label{tab:j1511v2}
\tablefoot{Columns: (1) Function used for modelling, (2) Magnitude, (3) Effective radius, (4) S\'{e}rsic index, (5) Axial ratio, (6) Position angle, (7) Additional notes on the sources.}
\end{table*}

\begin{figure*}
\centering
\adjustbox{valign=t}{\begin{minipage}{0.32\textwidth}
\centering
\includegraphics[width=1\textwidth]{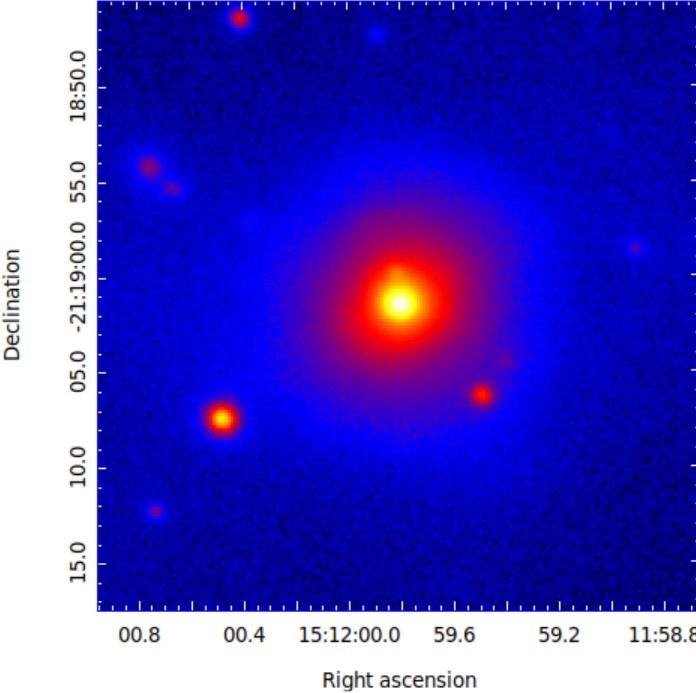}
\end{minipage}}
\adjustbox{valign=t}{\begin{minipage}{0.31\textwidth}
\centering
\includegraphics[width=0.95\textwidth]{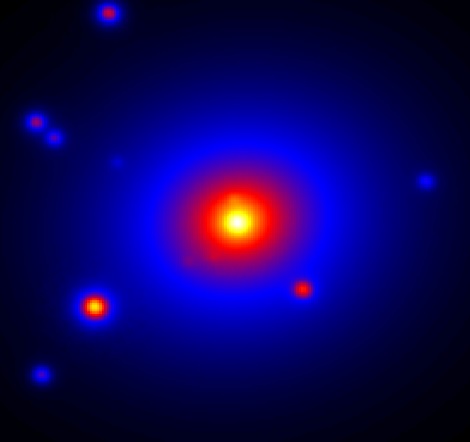}
\end{minipage}}
\adjustbox{valign=t}{\begin{minipage}{0.31\textwidth}
\centering
\includegraphics[width=0.95\textwidth]{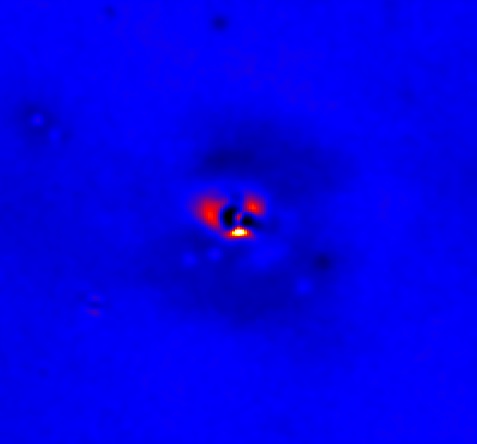}
\end{minipage}}
\hfill
    \caption{\textit{J}-band images of J1511-2119. The FoV is 31.8\arcsec    $/$ 26.5~kpc in all images. \emph{Left panel:} observed image, \emph{middle panel:} model image, and \emph{right panel:} residual image, smoothed over 3px.}  \label{fig:j1511v2}
\adjustbox{valign=t}{\begin{minipage}{0.49\textwidth}
\centering
\includegraphics[width=0.95\textwidth]{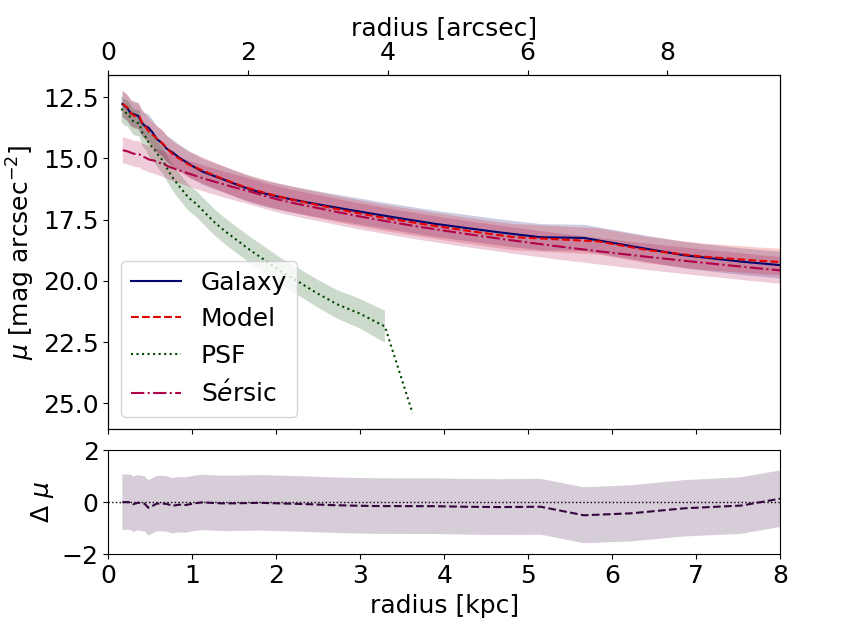}
\caption{Radial surface brightness profile plot of J1511-2119. The blue line represents the galaxy component, the dashed red line shows the model, the dotted green line marks the PSF, the dashed pink line shows the S\'{e}rsic 1 component, the final component, S\'{e}rsic 2, is shown with dotted light blue. The error of each component is shown by the shaded area surrounding each curve.}
\label{fig:j1511v2comps}
\end{minipage}}
\hfill
\adjustbox{valign=t}{\begin{minipage}{0.49\textwidth}
\centering
\includegraphics[width=0.95\textwidth]{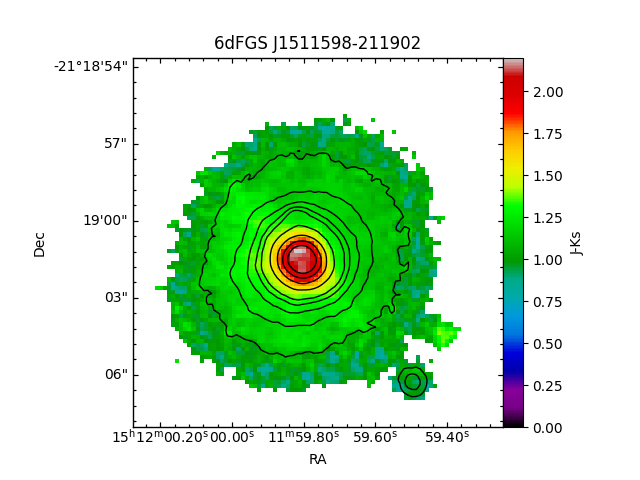}
\caption{\textit{$J$ - $Ks$} colour map of J1511-2119.}
\label{fig:j1511color}
\end{minipage}}
\hfil
\end{figure*}

This is one of the three jetted sources in this study. Its radio map can be found in \citet{2020chen}. We used a \textit{DAOPHOT} PSF model when fitting. There were ten nearby sources in our FoV, all of which we were able to fit with a single PSF with very little residuals. 

Our fit models a combination of a disk and a bulge (Table~\ref{tab:j1511v2}). The S\'{e}rsic index of 2.96 does not strongly support any specific morphology, even though indices this high can be found also in spiral galaxies \citep{2014elmegreen}. The size of the galaxy is typical for an NLS1 host. The excess residuals (Fig.~\ref{fig:j1511v2}) near the core imply the presence of unmodelled structures, such as leftover fragments of the bulge. The galaxy, model, and S\'{e}rsic curves in the radial surface brightness plot, Fig.~\ref{fig:j1511v2comps}, are all very similar to each other.

The host galaxy of J1511-2119 is disk-like, and based on the S\'{e}rsic index, an S0 galaxy. The colour map, shown in Fig.~\ref{fig:j1511color}, shows colours commonly seen in NLS1 galaxies, with the red most likely signifying dust extinction. The core of the galaxy is extremely red as the $J$-$Ks$ magnitude is $\sim$2. Beyond the core, the galaxy has very uniform colours.

\subsection{J1522-0644}

\begin{figure*}
\centering
\adjustbox{valign=t}{\begin{minipage}{0.31\textwidth}
\centering
\includegraphics[width=1\textwidth]{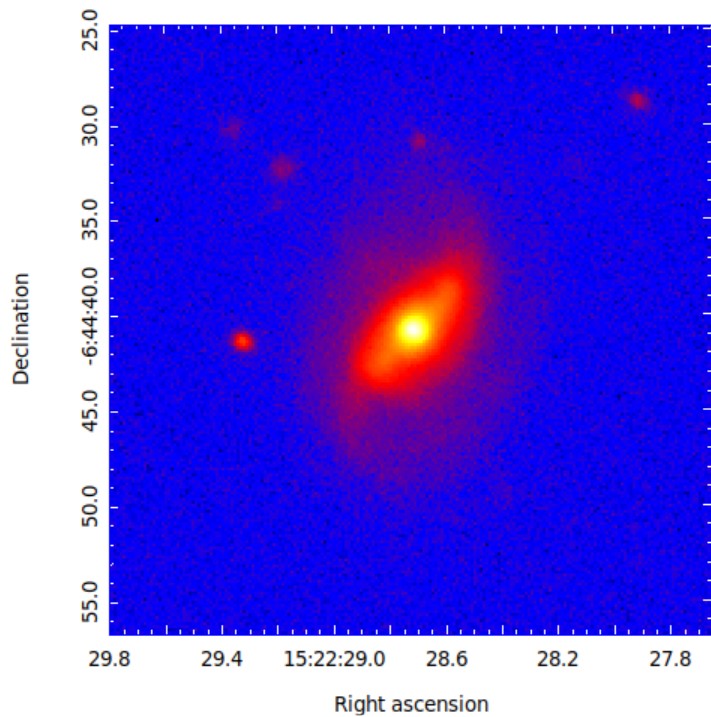}
\end{minipage}}
\adjustbox{valign=t}{\begin{minipage}{0.31\textwidth}
\centering
\includegraphics[width=0.95\textwidth]{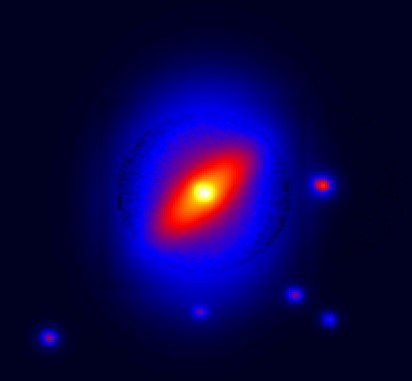}
\end{minipage}}
\adjustbox{valign=t}{\begin{minipage}{0.31\textwidth}
\centering
\includegraphics[width=0.95\textwidth]{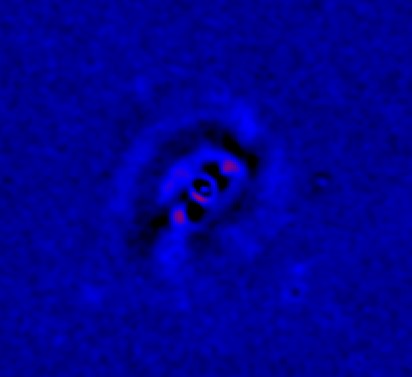}
\end{minipage}}
\hfill
    \caption{\textit{J}-band images of J1522-0644. The FoV is 31.8\arcsec    $/$ 47.7~kpc in all images. \emph{Left panel:} observed image, \emph{middle panel:} model image, and \emph{right panel:} residual image, smoothed over 3px.}  \label{fig:j1522}

\adjustbox{valign=t}{\begin{minipage}{0.49\textwidth}
\centering
\includegraphics[width=0.95\textwidth]{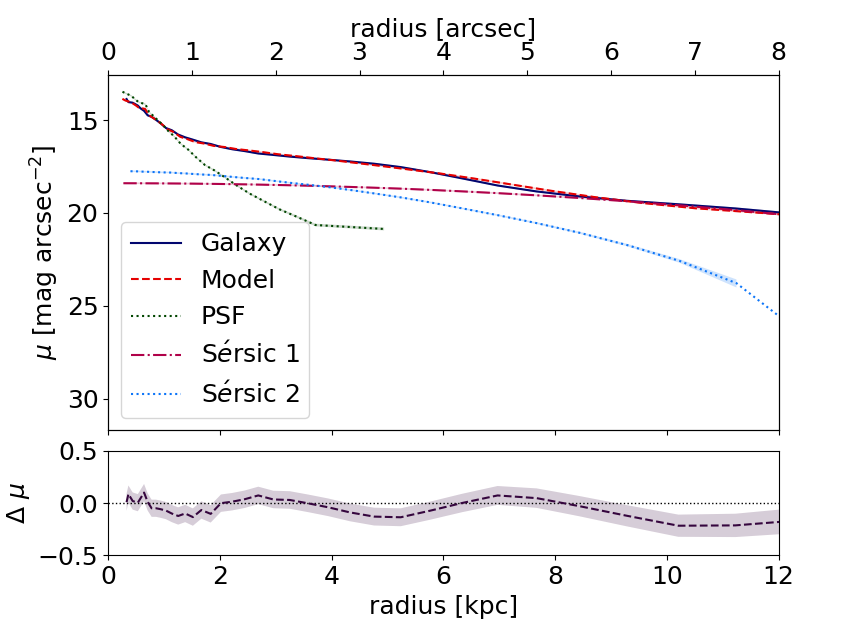}
\caption{Radial surface brightness profile plot of J1522-0644. The blue line represents the galaxy component, the dashed red line shows the model, the dotted green line marks the PSF, and finally the dashed pink line shows the S\'{e}rsic component. The error of each component is shown by the shaded area surrounding each curve. }
\label{fig:j1522comps}
\end{minipage}}
\hfill
\adjustbox{valign=t}{\begin{minipage}{0.49\textwidth}
\centering
\includegraphics[width=0.95\textwidth]{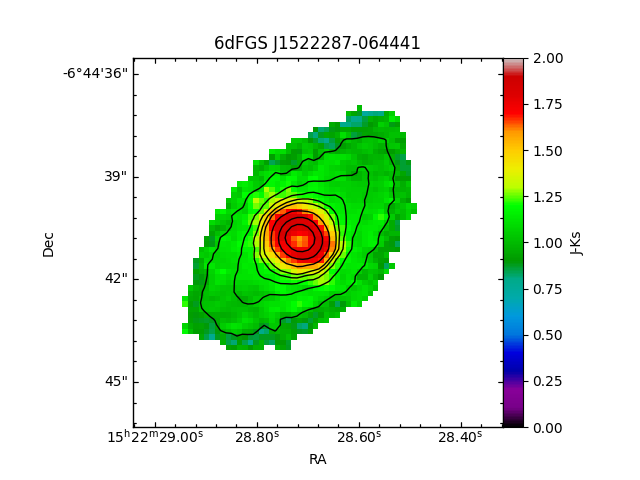}
\caption{\textit{$J$ - $Ks$} colour map of J1522-0644.}
\label{fig:j1522color}
\end{minipage}}
\hfill
\adjustbox{valign=t}{\begin{minipage}{0.49\textwidth}
\centering
\includegraphics[width=0.95\textwidth]{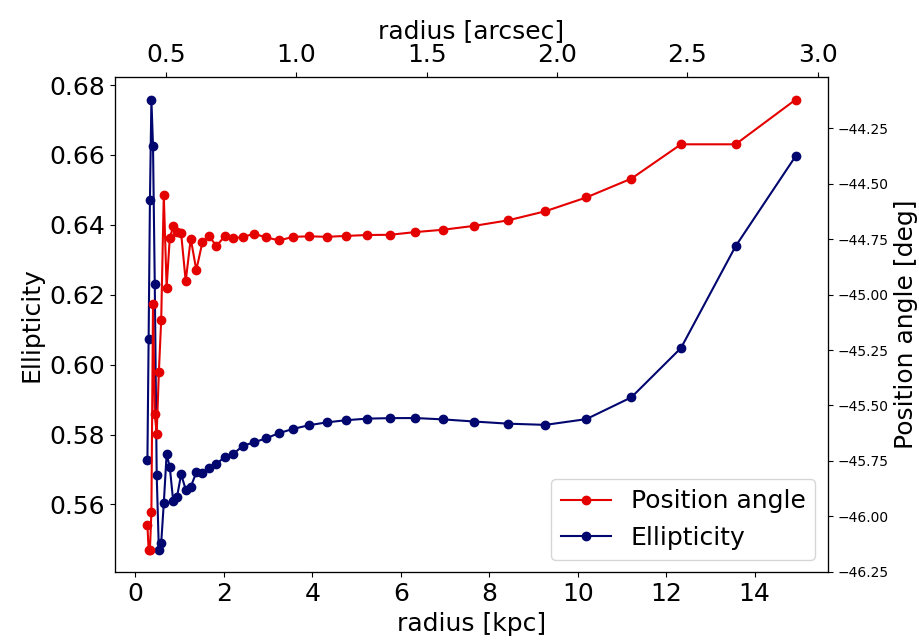}
\caption{Position angle and ellipticity of the S\'{e}rsic 1 function plotted against the major axis of the isophote of J1522-0644.}
\label{fig:j1522bar}
\end{minipage}}
\hfill
\end{figure*}

We successfully used the PSF model to fit the five nearby sources. The source, model, and residual images can be seen in Fig.~\ref{fig:j1522}. The best-fit parameters are given in Table~\ref{tab:j1522}. Along with a PSF, we used two S\'{e}rsic functions for fitting this source. Based on the effective radius of S\'{e}rsic 1 it is presumably fitting a disk. The S\'{e}rsic index is low for a disk, but not unreasonable. Based on its parameters, the second S\'{e}rsic function most likely models a bar. To further study this possibility, we plotted the PA and ellipticity over the radius, seen in Fig.~\ref{fig:j1522bar}. The PA remains steady throughout the plot. The ellipticity increases toward the higher radii. Based on this plot, it is difficult to confirm whether or not S\'{e}rsic 2 truly models a bar. The remaining residuals of the source are mostly remnants of the spiral arms. 

The $\chi^2_{\nu}$ is relatively high due to the remaining residuals, but the fit is satisfactory and the radial surface plot, Fig.~\ref{fig:j1522comps}, supports the goodness of the fit. The galaxy and the model curves match. The colour magnitude, seen in Fig.~\ref{fig:j1522color}, of the galaxy is as expected in an NLS1 galaxy. The $J$-$Ks$ magnitude near the core is approximately 1.75, and is thus very red. The redness of the core is most likely due to dust extinction.

\begin{table*}
\caption[]{Best-fit parameters of J1522-0644. $\chi^2_{\nu}$ = 1.257 $\substack{+0.03 \\ -0.03}$.}
\centering
\begin{tabular}{l l l l l l l}
\hline\hline
Function & Mag & $r_\text{e}$ & $n$ & Axial & PA & Notes \\
 & & (kpc) & & ratio & (\textdegree) & \\ 
\hline
PSF 1 & 15.55 $\substack{+0.05 \\ -0.05}$ & & & & & \\
S\'{e}rsic 1 & 14.76 $\substack{+0.01 \\ -0.00}$ & 7.72 $\substack{+0.02 \\ -0.02}$ & 0.48 $\substack{+0.01 \\ -0.01}$ & 0.76 $\substack{+0.00 \\ -0.00}$ & -5.97  $\substack{+0.02 \\ -0.02}$ & Disk \\
S\'{e}rsic 2 & 15.19 $\substack{+0.00 \\ -0.00}$ & 3.80 $\substack{+0.00 \\ -0.00}$ & 0.61 $\substack{+0.00 \\ -0.00}$ & 0.36 $\substack{+0.00 \\ -0.00}$ & -44.74  $\substack{+0.01 \\ -0.01}$ & Bar \\
%PSF 2 & 18.91 $\substack{+0.05 \\ -0.05}$ & & & & &  Nearby source \\
%PSF 3 & 20.27 $\substack{+0.06 \\ -0.06}$ & & & & &  Nearby source \\
%PSF 4 & 20.35 $\substack{+0.05 \\ -0.05}$ & & & & &  Nearby source \\
%PSF 5 & 20.05 $\substack{+0.06 \\ -0.06}$ & & & & &  Nearby source \\
%PSF 6 & 20.85 $\substack{+0.05 \\ -0.05}$ & & & & &  Nearby source \\
\hline   % centered "correctly"
\end{tabular}
\tablefoot{Columns: (1) Function used for modelling, (2) Magnitude, (3) Effective radius, (4) S\'{e}rsic index, (5) Axial ratio, (6) Position angle, (7) Additional notes on the sources.}
\label{tab:j1522}
\end{table*}

\subsection{J1615-0936}

We tried to model the PSF with \textit{DAOPHOT} and by directly using a nearby star. Based on our results, the \textit{DAOPHOT} PSF provided a better model. The unsatisfactory fit of the nearby source increases the $\chi^2_{\nu}$ of the fit.

\begin{figure*}
\centering
\adjustbox{valign=t}{\begin{minipage}{0.35\textwidth}
\centering
\includegraphics[width=1\textwidth]{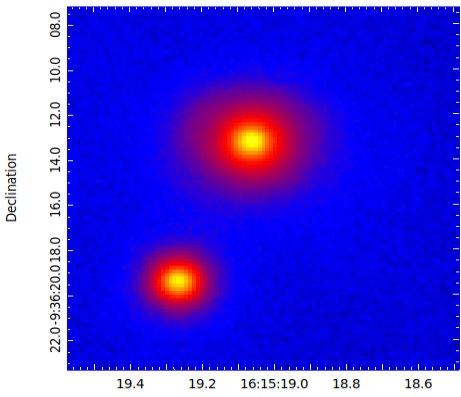}
\end{minipage}}
\adjustbox{valign=t}{\begin{minipage}{0.31\textwidth}
\centering
\includegraphics[width=0.95\textwidth]{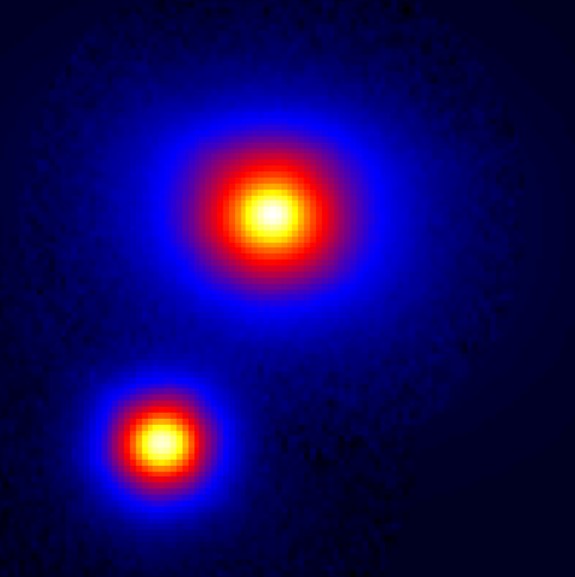}
\end{minipage}}
\adjustbox{valign=t}{\begin{minipage}{0.31\textwidth}
\centering
\includegraphics[width=0.95\textwidth]{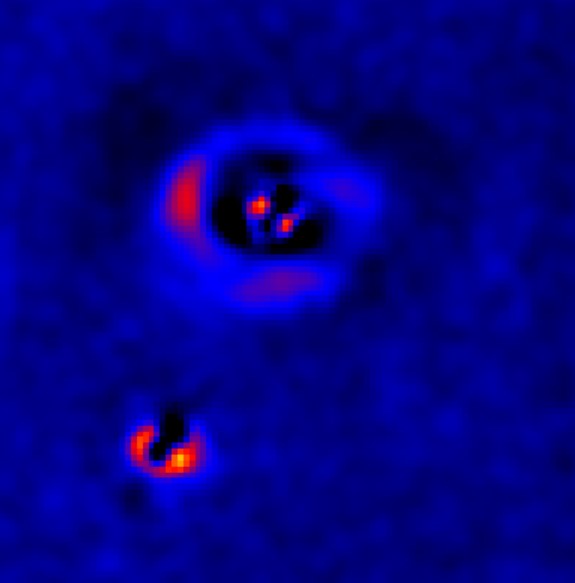}
\end{minipage}}
\hfill
    \caption{\textit{J}-band images of J1615-0936. The FoV is 15.9\arcsec    $/$ 18.8~kpc in all images. \emph{Left panel:} observed image, \emph{middle panel:} model image, and \emph{right panel:} residual image, smoothed over 3px.}  \label{fig:j1615}

\adjustbox{valign=t}{\begin{minipage}{0.49\textwidth}
\centering
\includegraphics[width=0.95\textwidth]{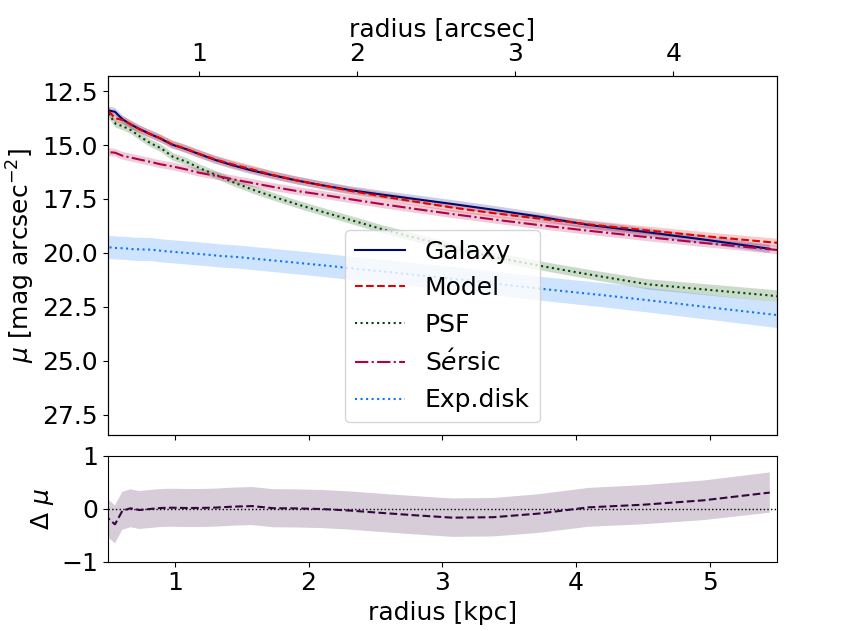}
\caption{Radial surface brightness profile plot of J1615-0936. The blue line represents the galaxy component, the dashed red line shows the model, the dotted green line marks the PSF, the dashed pink line shows the S\'{e}rsic component, the final component, the exponential disk, is shown with dotted light blue. The error of each component is shown by the shaded area surrounding each curve.}
\label{fig:j1615comps}
\end{minipage}}
\hfill
\adjustbox{valign=t}{\begin{minipage}{0.49\textwidth}
\centering
\includegraphics[width=0.95\textwidth]{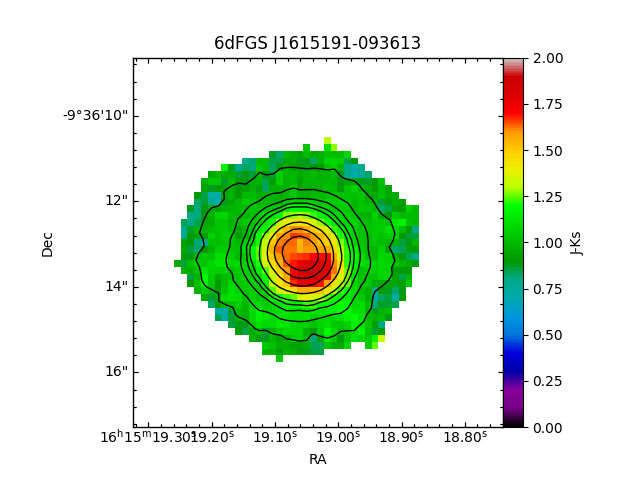}
\caption{\textit{$J$ - $Ks$} colour map of J1615-0936.}
\label{fig:j1615color}
\end{minipage}}
\hfill
\end{figure*}

We modelled the source with a PSF, a S\'{e}rsic function, and an exponential disk function. The best-fit parameters can be found in Table~\ref{tab:j1615}. Despite several attempts and different approaches, the $\chi^2_{\nu}$ remained quite high. Based on our best-fit results, the S\'{e}rsic represents a pseudo-bulge and the exponential disk function naturally fits a disk. There are some clear residuals left, as seen in Fig.~\ref{fig:j1615}. The residuals of the NLS1 are most likely caused by the bright AGN that was hard to model due to the lack of a bright enough PSF star. 

The $\chi^2_{\nu}$ of the fit is not ideal, however, it does not always imply that the fit is physically unreasonable \citep{2022dewsnap}. Indeed, based on the radial surface brightness plot, seen in Fig.~\ref{fig:j1615comps}, our model is accurate, as there is virtually no deviation between the model and galaxy curves, and the obtained components are physically reasonable. Based on our results, the host galaxy of J1615-0936 is disk-like, resembling a lenticular galaxy. This is supported by the axial ratio of the disk component being so small, 0.29 $\substack{+0.00 \\ -0.00}$.

The colour map of J1615-0936 is presented in Fig.~\ref{fig:j1615color}. Based on it, the core has a $J$-$Ks$ colour magnitude of roughly 1.5, which means that the core of the galaxy is very red. The rest of the galaxy presents uniform colours. Overall, the colours of the galaxy are as would be expected in an NLS1 galaxy.

\begin{table*}
\caption[]{Best-fit parameters of J1615-0936. $\chi^2_{\nu}$ = 2.553 $\substack{+0.02 \\ -0.01}$.}
\centering
\begin{tabular}{l l l l l l l}
\hline\hline
Function & Mag & $r_\text{e}$ & $n$ & Axial & PA & Notes \\
 & & (kpc) & & ratio & (\textdegree) & \\ 
\hline
PSF 1 & 14.25 $\substack{+0.03 \\ -0.03}$ & & & & & \\
S\'{e}rsic & 14.68 $\substack{+0.03 \\ -0.03}$ & 1.61 $\substack{+0.05 \\ -0.04}$ & 2.74 $\substack{+0.15 \\ -0.12}$ & 0.72 $\substack{+0.00 \\ -0.00}$ & 81.49  $\substack{+0.07 \\ -0.05}$ & Bulge \\
Exp.disk & 17.62 $\substack{+0.03 \\ -0.03}$ & 2.44 $\substack{+0.00 \\ -0.03}$ &  & 0.29 $\substack{+0.00 \\ -0.00}$ & 87.69  $\substack{+0.00 \\ -0.00}$ & Disk \\
%PSF 2 & 14.48 $\substack{+0.03 \\ -0.03}$ & & & & &  Nearby source \\
\hline   % centered "correctly"
\end{tabular}
\tablefoot{Columns: (1) Function used for modelling, (2) Magnitude, (3) Effective radius, (4) S\'{e}rsic index, (5) Axial ratio, (6) Position angle, (7) Additional notes on the sources.}
\label{tab:j1615}
\end{table*}

\subsection{J1628-0304}

\begin{figure*}
\centering
\adjustbox{valign=t}{\begin{minipage}{0.35\textwidth}
\centering
\includegraphics[width=1\textwidth]{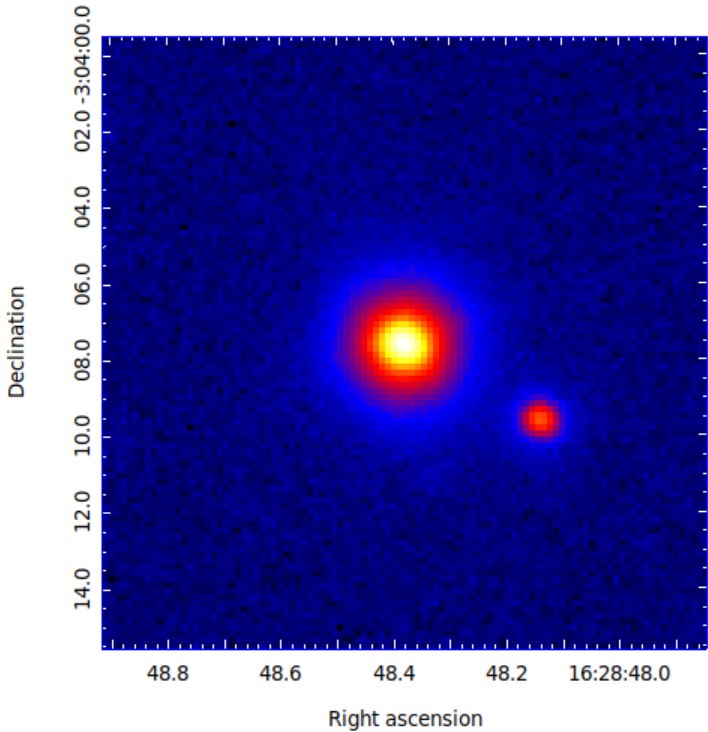}
\end{minipage}}
\adjustbox{valign=t}{\begin{minipage}{0.31\textwidth}
\centering
\includegraphics[width=0.95\textwidth]{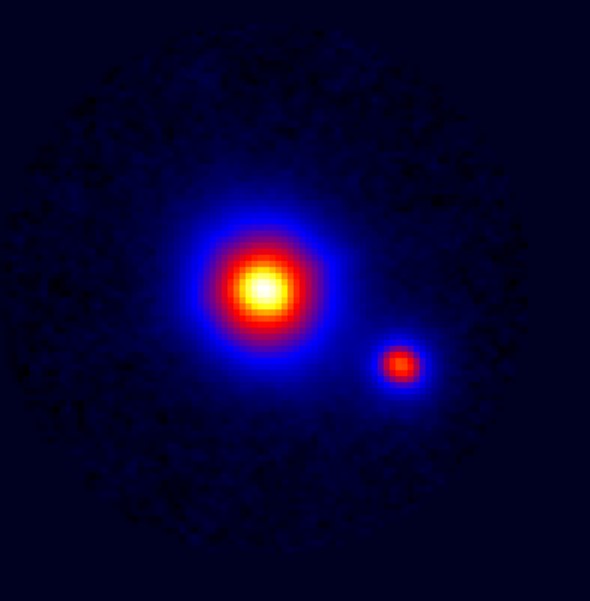}
\end{minipage}}
\adjustbox{valign=t}{\begin{minipage}{0.31\textwidth}
\centering
\includegraphics[width=0.95\textwidth]{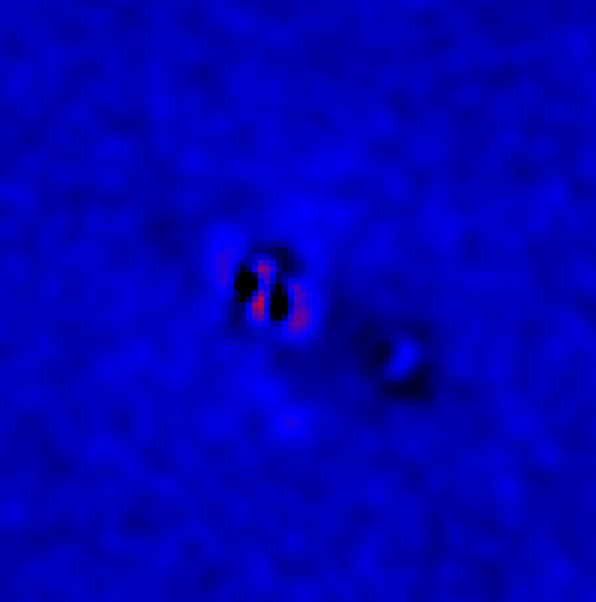}
\end{minipage}}
\hfill
    \caption{\textit{J}-band images of J1628-0304. The FoV is 15.9\arcsec    $/$ 26.2~kpc in all images. \emph{Left panel:} observed image, \emph{middle panel:} model image, and \emph{right panel:} residual image, smoothed over 3px.}  \label{fig:j1628}

\adjustbox{valign=t}{\begin{minipage}{0.49\textwidth}
\centering
\includegraphics[width=0.95\textwidth]{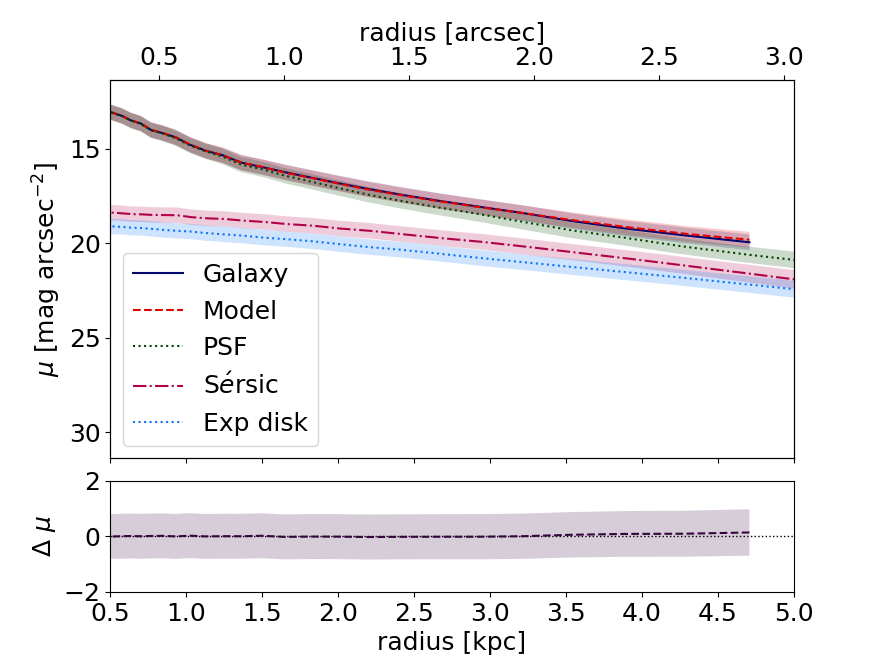}
\caption{Radial surface brightness profile plot of J1628-0304. The blue line represents the galaxy component, the dashed red line shows the model, the dotted green line marks the PSF, the dashed pink line shows the S\'{e}rsic component, the final component, the exponential disk component is shown with dotted light blue. The error of each component is shown by the shaded area surrounding each curve.}
\label{fig:j1628comps}
\end{minipage}}
\hfill
\adjustbox{valign=t}{\begin{minipage}{0.49\textwidth}
\centering
\includegraphics[width=0.95\textwidth]{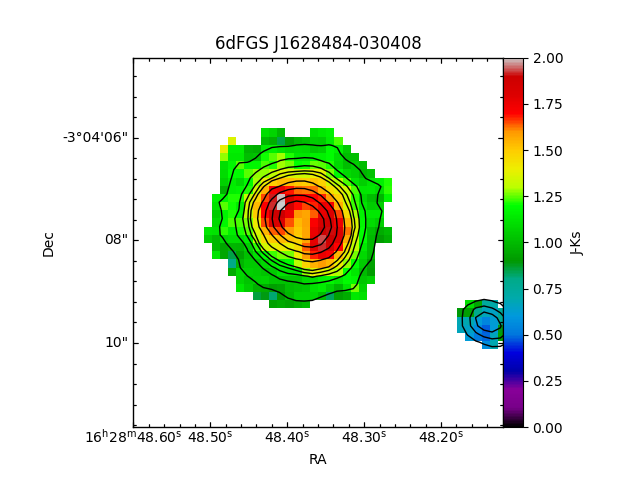}
\caption{\textit{$J$ - $Ks$} colour map of J1628-0304.}
\label{fig:j1628color}
\end{minipage}}
\hfill
\adjustbox{valign=t}{\begin{minipage}{0.49\textwidth}
\centering
\includegraphics[width=0.95\textwidth]{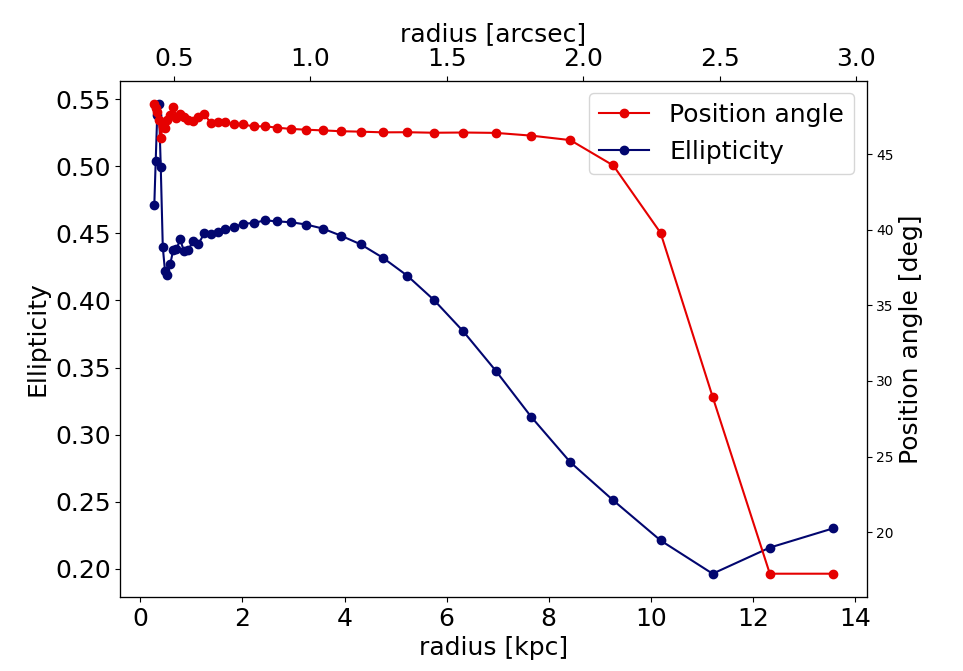}
\caption{Position angle and ellipticity of the S\'{e}rsic function plotted against the major axis of the isophote of J1628-0304.}
\label{fig:j1628bar}
\end{minipage}}
\hfill
\end{figure*}

We used a S\'{e}rsic function, an exponential disk function, and a PSF to fit the galaxy. The best-fit parameters are shown in Table~\ref{tab:j1628}. \textit{DAOPHOT} provided the best PSF model, but it is not perfect, as evidenced by the remaining residuals, seen in Fig.~\ref{fig:j1628}, near the core of the galaxy. The S\'{e}rsic function models a bar component. The galaxy is small, also proven by the small effective radius of the exponential disk component. We plotted the ellipticity and PA against the radius of the isophotes, seen in Fig.~\ref{fig:j1628bar}, to study the possible bar. In the plot the ellipticity can be seen increasing while the PA is more or less constant. The plot strongly suggests that the S\'{e}rsic function indeed models a bar component.

The $\chi^2_{\nu}$ for this source is quite high. However, there are virtually no high surface brightness residuals left. The remaining residuals are likely caused by the imperfect PSF model, which we could not improve despite several attempts with different methods. A major issue with the PSF modelling was the lack of bright enough stars in the FoV -- none of them are brighter than the AGN, and probably cannot model the wings of the PSF accurately. However, as discussed earlier, the $\chi^2_{\nu}$ value does not alone determine if the fit is physically justifiable and we want to emphasise the importance of taking into account also the obtained components, the residuals, and the comparison between the galaxy and and the model, in Fig.~\ref{fig:j1628comps}, when evaluating the fit. Based on all these aspects, the obtained results are adequate. The host galaxy of J1628-0304 is disk-like. 

The colour map can be seen in Fig.~\ref{fig:j1628color}. The colours of this galaxy are interesting as the reddest part is not at the core, but rather surrounds it. The structure could be due to a star forming ring or a minor merger \citep{2014iodice}. However, the resolution of our images does not allow a detailed view of the possible minor merger and due to this, the feature is most likely a star forming ring. Circumnuclear star forming rings have been frequently observed in NLS1 galaxies \citep{2022winkel}. Nonetheless, the core has colour magnitude of $\sim$1.5 and is thus very red.

\begin{table*}
\caption[]{Best-fit parameters of J1628-0304. $\chi^2_{\nu}$ = 3.183 $\substack{+0.05 \\ -0.05}$.}
\centering
\begin{tabular}{l l l l l l l}
\hline\hline
Function & Mag & $r_\text{e}$ & $n$ & Axial & PA & Notes \\
 & & (kpc) & & ratio & (\textdegree) & \\ 
\hline
PSF 1 & 14.37 $\substack{+0.16 \\ -0.17}$ & & & & & \\
S\'{e}rsic & 18.16 $\substack{+0.18 \\ -0.17}$ & 1.92 $\substack{+0.03 \\ -0.00}$ & 0.75 $\substack{+0.00 \\ -0.00}$ & 0.44 $\substack{+0.00 \\ -0.00}$ & 46.06  $\substack{+0.15 \\ -0.00}$ &  Bar   \\
Exp. disk & 17.45  $\substack{+0.16 \\ -0.17}$ & 2.00 $\substack{+0.01 \\ -0.02}$ & & 0.57 $\substack{+0.01 \\ -0.00}$   & 43.23  $\substack{+0.31 \\ -0.00}$ &  Disk  \\
%PSF 2 & 17.53 $\substack{+0.16 \\ -0.17}$ & & & & &  Nearby source \\
\hline   % centered "correctly"
\end{tabular}
\tablefoot{Columns: (1) Function used for modelling, (2) Magnitude, (3) Effective radius, (4) S\'{e}rsic index, (5) Axial ratio, (6) Position angle, (7) Additional notes on the sources.}
\label{tab:j1628}
\end{table*}

\subsection{J1646-1124}

\begin{figure*}
\centering
\adjustbox{valign=t}{\begin{minipage}{0.35\textwidth}
\centering
\includegraphics[width=1\textwidth]{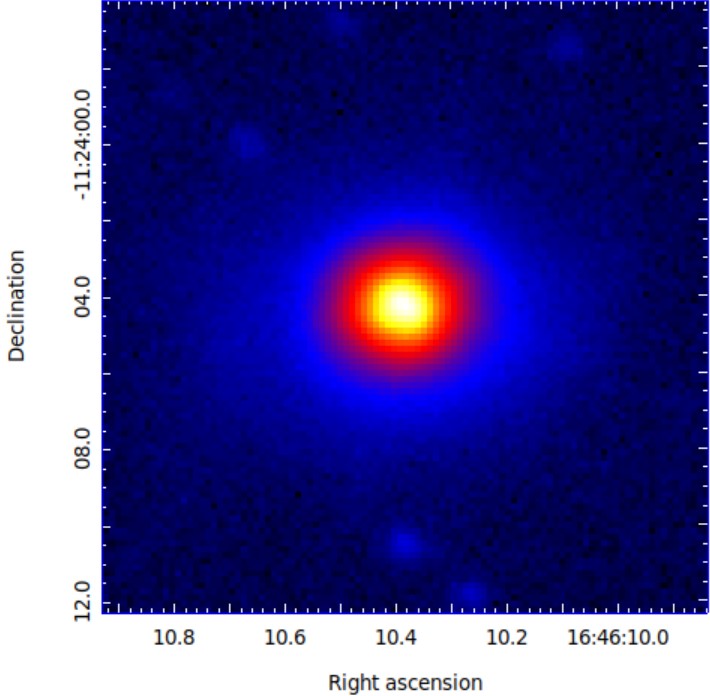}
\end{minipage}}
\adjustbox{valign=t}{\begin{minipage}{0.31\textwidth}
\centering
\includegraphics[width=0.95\textwidth]{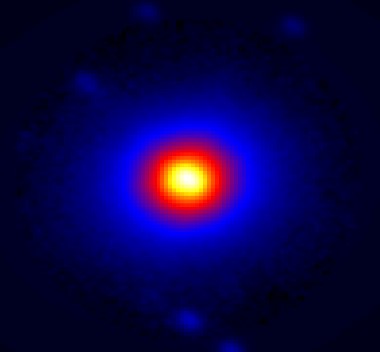}
\end{minipage}}
\adjustbox{valign=t}{\begin{minipage}{0.31\textwidth}
\centering
\includegraphics[width=0.95\textwidth]{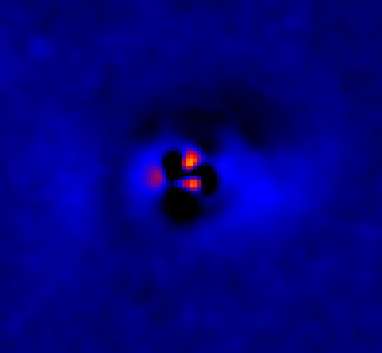}
\end{minipage}}
\hfill
    \caption{\textit{J}-band images of J1646-1124. The FoV is 15.9\arcsec    $/$ 21.5~kpc in all images. \emph{Left panel:} observed image, \emph{middle panel:} model image, and \emph{right panel:} residual image, smoothed over 3px.}  \label{fig:j1646}

\adjustbox{valign=t}{\begin{minipage}{0.49\textwidth}
\centering
\includegraphics[width=0.95\textwidth]{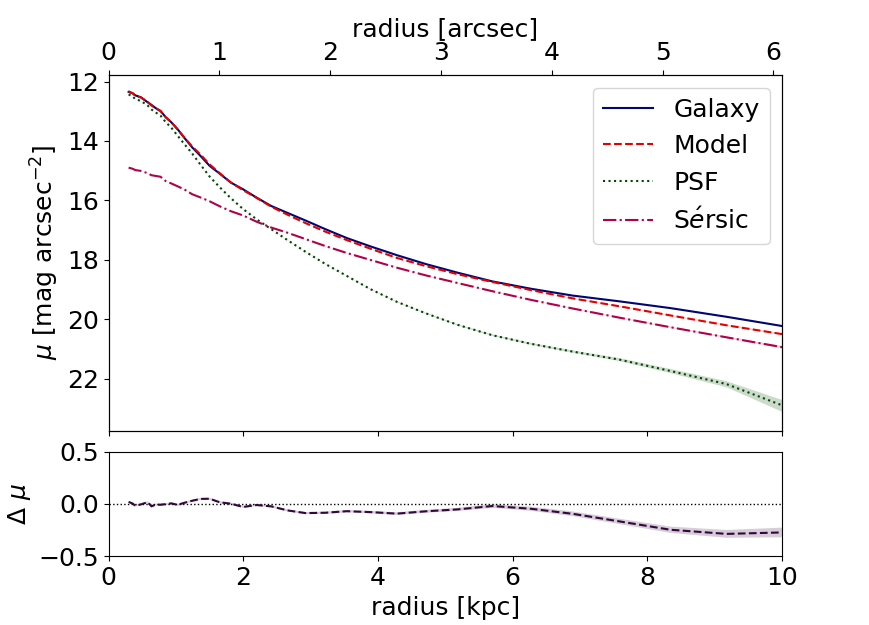}
\caption{Radial surface brightness profile plot of J1646-1124. The blue line represents the galaxy component, the dashed red line shows the model, the dotted green line marks the PSF, the dashed pink line shows the S\'{e}rsic 1 component, the final component, S\'{e}rsic 2, is shown with dotted light blue. The error of each component is shown by the shaded area surrounding each curve.}
\label{fig:j1646comps}
\end{minipage}}
\hfill
\adjustbox{valign=t}{\begin{minipage}{0.49\textwidth}
\centering
\includegraphics[width=0.95\textwidth]{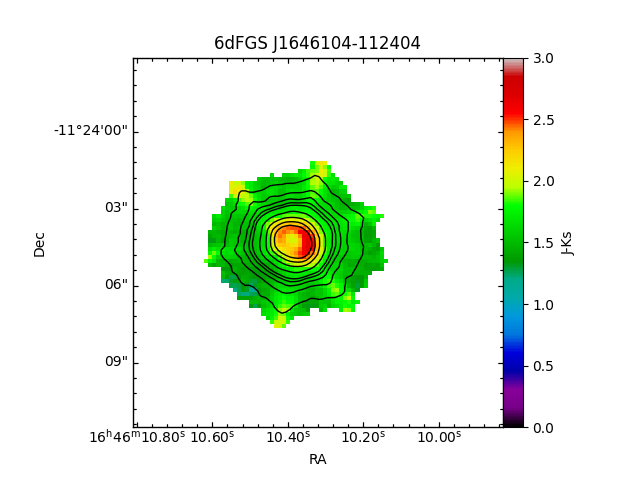}
\caption{\textit{$J$ - $Ks$} colour map of J1646-1124.}
\label{fig:j1646color}
\end{minipage}}
\hfill
\end{figure*}

This source has a high $\chi^2_{\nu}$ value (Table~\ref{tab:j1646}). Despite this, when studying the residuals, seen in Fig.~\ref{fig:j1646}, and the best-fit parameters, the fit is acceptable. A PSF was fitted on the five nearby sources as well as on the galaxy. In addition, the galaxy required one S\'{e}rsic function, with the function most likely modelling a bulge. The S\'{e}rsic index is high for a spiral galaxy, though still within the range of accepted values for them. We attempted fitting additional components, but could not achieve a physically justifiable fit. Our issues mostly stem from the AGN being so bright. A fit with three S\'{e}rsic functions yielded a better $\chi^2_{\nu}$ (~2), but one of the components consistently remained unresolved, likely trying to fit parts of the point-like AGN. The resolved components modelled a bulge and a disk component. The alternative-fit results can be seen in Table~\ref{tab:j1646alt}.

Some residuals still remain near the galaxy core, and are most likely due to the issues with the PSF model. There is indication of a spiral-like structure surrounding the galaxy, starting from its west side, and bending anti-clockwise toward east. However, without further proof, the assessment of the residuals remains a speculation. Due to the $\chi^2_{\nu}$ value, the residuals, and the best-fit parameters, we do not deem it possible to irrefutably determine the host galaxy of J1646-1124. However, there is evidence, for example, in the alternative fit, that J1646-1124 is disk-like, and based on the parameters, an S0. 

The radial surface plot, seen in Fig.~\ref{fig:j1646comps}, shows that despite the difficulties in the modelling, the model fits the galaxy very well up to the radius of 6~kpc. The colour map of the source can be seen in Fig.~\ref{fig:j1646color}. The colours are as expected for an NLS1 galaxy, with the core being very red ($\sim$2.2). The redness is most likely due to dust extinction. Some slight saturation, resulting in the polygonal shape, due to the $Ks$-image, can be seen in the colour map.

\begin{table*}
\caption[]{Best-fit parameters of J1646-1124. $\chi^2_{\nu}$ = 3.221 $\substack{+0.02 \\ -0.01}$.}
\centering
\begin{tabular}{l l l l l l l}
\hline\hline
Function & Mag & $r_\text{e}$ & $n$ & Axial & PA & Notes \\
 & & (kpc) & & ratio & (\textdegree) & \\ 
\hline
PSF 1 & 13.76 $\substack{+0.00 \\ -0.00}$ & & & & & \\
S\'{e}rsic & 14.63 $\substack{+0.05 \\ -0.06}$ & 1.65 $\substack{+0.03 \\ -0.03}$ & 2.68 $\substack{+0.15 \\ -0.12}$ & 0.81 $\substack{+0.00 \\ -0.00}$ & -72.19  $\substack{+0.03 \\ -0.05}$ & Bulge \\
%PSF 2 & 20.7 $\substack{+0.00 \\ -0.00}$ & & & & &  Nearby source \\
%PSF 3 & 21.42 $\substack{+0.01 \\ -0.00}$ & & & & &  Nearby source \\
%PSF 4 & 21.52 $\substack{+0.00 \\ -0.02}$ & & & & &  Nearby source \\
%PSF 5 & 21.51 $\substack{+0.00 \\ -0.00}$ & & & & &  Nearby source \\
%PSF 6 & 20.44 $\substack{+0.00 \\ -0.00}$ & & & & &  Nearby source \\
\hline   % centered "correctly"
\end{tabular}
\label{tab:j1646}
\tablefoot{Columns: (1) Function used for modelling, (2) Magnitude, (3) Effective radius, (4) S\'{e}rsic index, (5) Axial ratio, (6) Position angle, (7) Additional notes on the sources.}
\end{table*}

\begin{table*}
\caption[]{Alternative-fit parameters of J1646-1124. $\chi^2_{\nu}$ = 2.229 $\substack{+0.03 \\ -0.04}$.}
\centering
\begin{tabular}{l l l l l l l}
\hline\hline
Function & Mag & $r_\text{e}$ & $n$ & Axial & PA & Notes \\
 & & (kpc) & & ratio & (\textdegree) & \\ 
\hline
PSF 1 & 13.83 $\substack{+0.00 \\ -0.00}$ & & & & & \\
S\'{e}rsic 1 & 15.94 $\substack{+0.05 \\ -0.06}$ & 1.88 $\substack{+0.02 \\ -0.04}$ & 1.38 $\substack{+0.04 \\ -0.21}$ & 0.72 $\substack{+0.00 \\ -0.00}$ & 66.76  $\substack{+0.03 \\ -0.05}$ & Disk \\
S\'{e}rsic 2 & 15.78 $\substack{+0.04 \\ -0.03}$ & 0.37 $\substack{+0.02 \\ -0.03}$ & 0.34 $\substack{+0.04 \\ -0.02}$ & 0.17 $\substack{+0.00 \\ -0.00}$ & 13.43  $\substack{+0.08 \\ -0.05}$ &  \\
S\'{e}rsic 3 & 15.23 $\substack{+0.05 \\ -0.06}$ & 3.11 $\substack{+0.24 \\ -0.13}$ & 2.52 $\substack{+0.11 \\ -0.12}$ & 0.70 $\substack{+0.00 \\ -0.00}$ & -89.6  $\substack{+0.01 \\ -0.05}$ & Bulge \\
%PSF 2 & 20.52 $\substack{+0.00 \\ -0.00}$ & & & & &  Nearby source \\
%PSF 3 & 21.23 $\substack{+0.01 \\ -0.00}$ & & & & &  Nearby source \\
%PSF 4 & 21.12 $\substack{+0.00 \\ -0.02}$ & & & & &  Nearby source \\
%PSF 5 & 21.05 $\substack{+0.00 \\ -0.00}$ & & & & &  Nearby source \\
%PSF 6 & 20.27 $\substack{+0.00 \\ -0.00}$ & & & & &  Nearby source \\
\hline   % centered "correctly"
\end{tabular}
\label{tab:j1646alt}
\tablefoot{Columns: (1) Function used for modelling, (2) Magnitude, (3) Effective radius, (4) S\'{e}rsic index, (5) Axial ratio, (6) Position angle, (7) Additional notes on the sources.}
\end{table*}

\section{Discussion}  
\label{discussion}

%Our findings
We were able to model six out of the seven sources in our sample. The only source we could not model was J0952-0136. A summary with the final verdicts of each source can be found in Table~\ref{tab:summary}. Two of our modelled sources, J1511-2119 and J1615-0936, have been speculated by \citet{2020chen} to be jetted based on their radio spectra. All six of our sources have disk-like hosts. The only uncertain host morphology is for J1646-1124, however, our results still suggest the possibility of disk-like morphology. Three of our sources show evidence for being in lenticular galaxies. However, it is unclear if the morphology truly is lenticular or if the results are caused by the AGN being too bright to have a reliable PSF model from the FoV. We were able to model a disk component in five of our sources, which naturally indicates disk-like morphology of the galaxy. A bar component can be found in two sources, J1522-0644 and J1628-0304. Based on the PA and ellipticity versus the major axis of the isophote plots, J1628-0304 is very likely to host a bar. We were able to model a bulge component in four sources. The modelled bulges have a slightly high S\'{e}rsic index to be considered pseudo-bulges in disk-like galaxies. The commonly used limit for distinguishing between pseudo- and classical bulges is $n = 2$, where indices below 2 are associated with pseudo-bulges and indices above with classical bulges \citep{2011orban}. However, the sources that have bulges with S\'{e}rsic indices above 2 are the sources we suspect of being S0s and it has been suggested that S0/a sources can have pseudo-bulges with S\'{e}rsic index $\approx 2.4$ with a standard deviation of 0.66 \citep{2004kormendy}. With this in mind, the modelled bulges are most-likely pseudo-bulges, which is in agreement with what is seen in disk-like galaxies.

%Comparing to previous studies
Our findings are in alignment with the majority of previous studies regarding the host galaxy morphology of jetted and non-jetted NLS1 galaxies \citep[e.g.][]{2003crenshaw, 2007ohta1,2008anton, 2011orban, 2014leontavares, 2016kotilainen, 2017olguiglesias, 2018jarvela, 2019berton, 2020olguiglesias, 2021hamilton, 2022vietri, 2022varglund}. Only two jetted NLS1 galaxies have so far been identified to be located in elliptical galaxies \citep{2017dammando, 2018dammando}. However, the morphology of FBQS J1644+2619 being elliptical is up for debate as the source has also been suggested to reside in a barred lenticular galaxy. 

Only a few studies on NLS1 galaxy stellar masses exist, and we were able to find a stellar mass estimate for three of our sources: J0952-0136 ($M_{\text{star}} \sim 1.5 \times 10^{9} M_{\sun}$) \citep{2021pan}, J0622-2317 ($M_{\text{star}} \sim 1.3 \times 10^{10} M_{\sun}$) \citep{2016saulder}, and J1511-2119 ($M_{\text{star}} \sim 1.3 \times 10^{11} M_{\sun}$) \citep{2021koss}. Two of these sources have a low stellar mass, while J1511-2119 has a high stellar mass. The high stellar mass is not surprising as the galaxy is most likely an S0. The stellar mass of J0952-0136 being low is interesting as it is a jetted source, however, as we do not have information of the stellar mass of other jetted NLS1 galaxies, no further conclusions can be drawn. Based on the relation of the stellar mass, and the effective radius and S\'{e}rsic index of the NLS1 galaxies of this paper, the stellar mass of these sources is most likely low \citep{2020sanchez}.

% Interaction and mergers
When studying the evolution of AGN, the nearby environment, both group- and galaxy-scale, is a crucial factor to take into account. This is due to the nearby environment being capable of reshaping the galaxy as well as the galaxy dynamics. It has been hypothesised that jetted NLS1 galaxies and interaction/mergers might be linked \citep[e.g.][]{2017jarvela, 2020olguiglesias, 2021berton}. However, there is no clear consensus on the topic yet. None of the sources in this paper present clear mergers or interaction. This could be for a variety of reasons, one of the possible reasons being the lack of resolution for seeing minor mergers. It has been shown before that minor mergers are capable of triggering nuclear activity, an example of this is IRAS 17020+4544. Although this source is a jetted NLS1 galaxy, there is no clear evidence of mergers in optical/NIR imaging, but there is clear evidence of a small companion galaxy based on molecular gas observations \citep{2021salome, 2022jarvela}. 

It has also been speculated that $\gamma$-ray-emitting NLS1 galaxies have experienced more minor mergers than non-$\gamma$-NLS1 galaxies with significant radio emission \citep{2020olguiglesias}. The same study also stated that the $\gamma$-NLS1 galaxies have bulge colours that are significantly different from their hosts with red NIR colour, $J - K$ $\sim$ 2, while the non-$\gamma$-ray NLS1 galaxies have bulge colours that are similar to the rest of the galaxy. The red nuclei is, according to the authors, most likely due to either the bulge showing enhanced star formation or the AGN being dust-embedded as a consequence of a past minor merger. Contradictory to the results in \citet{2020olguiglesias}, all but one of the sources in this paper showcase the nucleus being significantly more red than the rest of the galaxy. The only exception to this is J0622-2317, however, it is possible that the colouration difference in this source is due to the spiral arms beginning directly from the core, thus "hiding" the core under the spiral arms. In case the speculation by \citet{2020olguiglesias} is indeed correct, it is entirely possible that the colour of the core could be a hint of previous minor mergers also in our sources. However, unlike in \citet{2020olguiglesias}, the nuclei of our sources do not present any clear difference in colours between jetted and non-jetted NLS1 galaxies. Based on the current understanding, it is impossible to say whether or not mergers are needed for triggering powerful relativistic jets in NLS1 galaxies.

\begin{table*}
\caption[]{Summary of the results. Columns: (1) Source name, (2) Probable morphology of the host galaxy, (3) Components of the model, (4) Presence of a relativistic jet based on the Metsähovi 37 GHz detections, (5) Large-scale environment parameter, (6) Additional notes.}
\centering
\begin{tabular}{l l l l l l}
\hline\hline
Source              & Morphology & Components     & Jetted &  Notes   \\
                    &            &                &        &            \\ \hline
6dFGS J0622335-231742 & Disk-like & Disk and bulge        & No &   \\
6dFGS J0952191-013644 & Unclear   &                       & Yes & Saturated image  \\
6dFGS J1511598-211902 & Disk-like & Disk and bulge        & Yes  & Possibly S0 \\
6dFGS J1522287-064441 & Disk-like & Disk and bar          & No  &  \\
6dFGS J1615191-093613 & Disk-like & Disk and bulge        & Yes & Possibly S0  \\
6dFGS J1628484-030408 & Disk-like & Disk and bar          & No  &  \\
6dFGS J1646104-112404 & Disk-like & Bulge                 & No  & Slightly saturated colour map, possibly S0 \\
\\\hline
\end{tabular}
\label{tab:summary}
\end{table*}

\section{Conclusions}
\label{conclusions}

We have studied six NLS1 galaxies that have been observed in $J-$ and $Ks$-bands. The observations were carried out with the FourStar NIR instrument. The goal of our study was to investigate what morphology the host galaxies of our sources have. We also created colour maps for all of our sources. These maps were created to help study the properties of the core of the source. Furthermore, we also wanted to study possible interaction in our sources. The main findings of our study are:

\begin{enumerate}
    \item All of our sources are most likely located in disk-like galaxies, including the possibly jetted NLS1 galaxies. 
    \item The colour maps show that the core of NLS1 galaxies strongly tends toward the red-side, with no significant difference between the cores of jetted and non-jetted NLS1 galaxies.
    \item The redness of the core is most likely due to either dust extinction or due to previous minor mergers.
    \end{enumerate}

With over 100 studied hosts of NLS1 galaxies, jetted and non-jetted, the host galaxies of all types of NLS1 nuclei can be stated to be predominantly disk-like. What is still mostly missing is the information regarding the host galaxies of NLS1 sources at high redshifts, and more observations will be needed to determine if their properties are similar to their low-redshift counterparts. Suitable instruments for these studies would be, for example, the Very Large Telescope as well as the JWST. Multi-wavelength observations are crucial for understanding the role of mergers and interaction in jetted NLS1 galaxies. For this, observations with the Atacama Large Millimeter Array at sub-mm would be a key factor. NLS1 galaxies offer a unique opportunity to study the interplay of different phenomena, such as jets, outflows, and star formation, at an early stage of AGN evolution, and estimate the impact an active nucleus has on the galaxy evolution. 

\begin{acknowledgements}
IRAF is distributed by the National Optical Astronomy Observatories, which are operated by the Association of Universities for Research in Astronomy, Inc., under cooperative agreement with FourStar, an infrared camera, part of the Magellan 6.5~m telescopes, at the Las Campanas Observatory in Chile. This research has made use of GALFIT. This publication made use of the SIMBAD database, operated at CDS, Strasbourg, France. This research has made use of "Aladin sky atlas" developed at CDS, Strasbourg Observatory, France. This research has made use of the NASA/IPAC Extragalactic Database (NED), which is operated by the Jet Propulsion Laboratory, California Institute of Technology, under contract with the National Aeronautics and Space Administration. This research made use of Astropy,\footnote{http://www.astropy.org} a community-developed core Python package for Astronomy \citep{astropy:2013, astropy:2018}. This publication makes use of data products from the Two Micron All Sky Survey, which is a joint project of the University of Massachusetts and the Infrared Processing and Analysis Center/California Institute of Technology, funded by the National Aeronautics and Space Administration and the National Science Foundation. IV would like to thank both the Magnus Ehrnrooth Foundation and the Vilho, Yrjö and Kalle Väisälä Foundation of the Finnish Academy of Science and Letters for their support. MB and EC are ESO fellows. 

\end{acknowledgements}

\bibliographystyle{aa}
\bibliography{articles.bib}

\end{document}